ORIGINAL PAPER

# A formalism for modelling traction forces and cell shape evolution during cell migration in various biomedical processes

Q. Peng[1,2] · F. J. Vermolen[1,2] · D. Weihs[3]

Received: 9 December 2020 / Accepted: 31 March 2021

## Abstract
The phenomenological model for cell shape deformation and cell migration Chen (BMM 17:1429–1450, 2018), Vermolen and Gefen (BMM 12:301–323, 2012), is extended with the incorporation of cell traction forces and the evolution of cell equilibrium shapes as a result of cell differentiation. Plastic deformations of the extracellular matrix are modelled using morphoelasticity theory. The resulting partial differential differential equations are solved by the use of the finite element method. The paper treats various biological scenarios that entail cell migration and cell shape evolution. The experimental observations in Mak et al. (LC 13:340–348, 2013), where transmigration of cancer cells through narrow apertures is studied, are reproduced using a Monte Carlo framework.

## 1 Introduction

Cells may attain various shapes and sizes, for example, stem cells can differentiate and adopt the shape and functionality of many different cell types in our body: fan-like keratocytes, hand-shaped nerve growth cones and spindle-shaped fibroblasts Mogilner and Keren (2009), Robey (2017). It has been recognized that cell geometry influences cellular activities like cell growth and death, cell mobility and adhesion to the direct environment Barnhart et al. (2011), Keren et al. (2008), Massalha and Weihs (2016), Mogilner and Keren (2009), Saeed and Weihs (2019). The shape of a motile cell is determined by its boundaries, which dynamically vary with a local balance between retraction and protrusion Ebata et al. (2018). There are multiple constituent elements affecting the cell shape, for instance, the cytoskeleton and the cell-substrate adhesions, which have been studied in depth in the past years. However, it is still a great challenge to understand the mechanisms that determine the global cell morphology in the context of its function Keren et al. (2008), Mogilner and Keren (2009).

Signalling molecules play an important role in cell migration and cell shape. During wound healing, chemotaxis is one of the most important cues for migration of immune cells and fibroblasts in inflammatory and proliferative phases Koppenol (2017), Cumming et al. (2009), Enoch and Leaper (2008), Peng and Vermolen (2020). Metastasis of cancer cells can be induced by nutrients and oxygen, since tumour growth requires an adequate supply of oxygen and nutrients. Under most pathological circumstances, oxygen and nutrients are supplied though the local blood vasculature Wek and Staschke (2010), Siemann and Horsman (2015). Commonly, signalling molecules are activated at the plasma membrane and de-activated in the cytoplasm. On the other hand, the concentration of signalling molecules determines the cytoskeletal dynamics Mogilner and Keren (2009).

In wound healing, cells migrate and change shape in both the epidermis and the dermis layers. Re-epithelialization is the most essential part for the skin to re-establish its barrier function Safferling et al. (2013), Singer and Clark (1999), Friedl and Gilmour (2009). However, the mechanisms of re-epithelialization are poorly understood. In the early stage of

✉ Q. Peng
Q.Peng-1@tudelft.nl

1 Delft Institute of Applied Mathematics, Delft University of Technology, Mekelweg 4, 2628 CD Delft, The Netherlands

2 Computational Mathematics Group, Discipline group Mathematics and statistics, Faculty of Science, Hasselt University, Campus Diepenbeek, Agoralaan Gebouw D, 3590 BE Diepenbeek, Belgium

3 Faculty of Biomedical Engineering, Technion-Israel Institute of Technology, 3200003 Haifa, Israel

the epidermis closure in a wound, the basement membrane between the epidermis and dermis extends slightly over the ends of the incised dermis, creating an "extension membrane" (or the so-called epidermal tongue) Rittié (2016). The mechanism of the occurrence of the epidermal tongue is still unclear. A possible explanation is that the suprabasal cells (which lie upon the layer of basal cells) form the tongue by migrating over the leading basal cells and dedifferentiating to basal cells (which are adhered to the basement membrane between the epidermis and dermis) to form new leaders Safferling et al. (2013), Rittié (2016), Vermolen and Javierre (2011), Rousselle et al. (2019). When epidermal epithelial cells are "crawling" and "climbing up" to reestablish the epidermis, they elongate and flatten Safferling et al. (2013). In the dermis, it has been widely documented that the differentiation of fibroblasts is one of the key events during wound healing. Differentiation changes the spindle-shaped fibroblast to dendritic-shaped myofibroblasts. Subsequently, cells' mechanobiology is modified considerably as well. The differentiated myofibroblasts exert much larger forces on the extracellular matrix (ECM) than fibroblasts Peng and Vermolen (2020). Excessive numbers of myofibroblasts will result in contractures, which are morbid and pathological macro-scale contractions. Usually, contractures concur with disabilities and dysfunction and have a grave impact on patients' daily life.

Cancer metastasis has been reported as the main reason of death in cancer patients Massalha and Weihs (2016). During the migration of a cancer cell to its destination, especially migrating through a narrow and stiff cavity, it has to deform to adapt to the obstacles. More invasive cancer cells appear be more pliable and dynamic both internally Gal and Weihs (2012) and externally Guck et al. (2005), Cross et al. (2007), Swaminathan et al. (2011) and thus able to adjust their cytoskeleton and morphology, which might provide a possible diagnosis for cancer. In addition to that, cancer cells are observed to apply a significantly larger traction force on the substrate, compared to benign cells Massalha and Weihs (2016), yet the specific mechanisms that induce these increased forces are still poorly understood.

Mathematical modelling has been proven to be an important tool to have a deeper insight into many biological processes that are potentially difficult to control in experiments, for example, wound healing and tumour growth. Depending on the scale of the observed domain, continuum models and agent-based models are widely used. Continuum models have the advantage of modelling a larger scale; however, the model neglects the individual cellular activity and cells are not tracked Vermolen and Gefen (2012). Agent-based model is suitable to model cellular activities of every cell, for instance, cell migration and cell deformation. Hence, an agent-based model is selected in this manuscript, and this work is an extension of Chen et al. (2018) and Vermolen and Gefen (2012). In Chen et al. (2018), a model of the deformation of both the cell and the nucleus is developed. Furthermore, a parameter sensitivity analysis is carried out on the basis of Monte Carlo simulations. However, the study in Chen et al. (2018) does not consider the traction forces applied by the cell and the impact on the substrates. Compared to the work of Vermolen and Gefen (2012), we use finite-element methods to solve all the partial differential equations, rather than Green's functions. Therefore, a more precise solution is delivered. Furthermore, we implement a more intricate approach to model the traction forces applied by cells in various applications. In addition to circular projections of cells in Chen et al. (2018) and Vermolen and Gefen (2012), we model elliptic and hypocycloid-shaped cells in this manuscript.

This manuscript is structured as follows: Sect. 2 explains the agent-based model of cell migration, in the form of a set of partial differential equations. Possible applications of this model and the corresponding numerical results are exhibited in Sect. 3. Finally, conclusions are shown in Sect. 4.

## 2 Mathematical modelling

In this manuscript, the phenomenological model of cell deformation is extended from the work in Chen et al. (2018), Vermolen and Gefen (2012), in particular, in two dimensions. With essential biological assumptions and simplifications, the model mainly describes the impact of extracellular components on the cell deformation and displacement. Subsequently, more applications can be developed, for instance, cell differentiation and cell repulsion. Different from the work in Zhao et al. (2020), where they also model the dynamics of intercellular adhesion by connecting a certain series of points inside the cell with elastic springs, the model in this manuscript neglects the intracellular environment; hence, it is not capable to present the Poisson's effect of the cell.

The cell membrane is split into finite line segments by the nodal points, and the centre position of cell is determined by the mean of all the positions of the nodal points. The equilibrium shape of the cell is kept by a collection of springs, which connects each nodal point on the cell membrane to the centre of the cell, respectively; see Fig. 1. For each nodal point, the displacement is determined by various mechanisms of directed motion and random turning Ebata et al. (2018), which will be discussed in details in the following contents. Regarding different applications of this model, there will be some model adjustments.

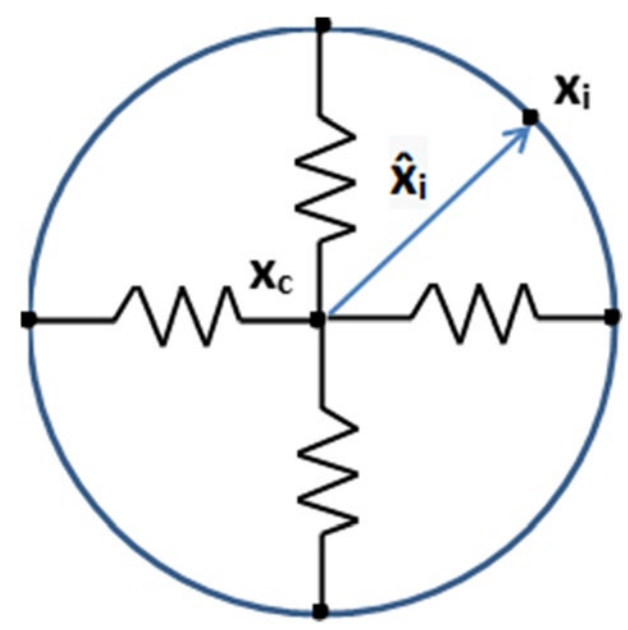

**Fig. 1** A schematic of the distribution of the nodal points on the cell membrane. In our model, we assume the equilibrium shape of the cell is maintained by a collection of springs

### 2.1 Concentration of generic signal

We assume that cell migration is mainly driven by chemotaxis (or mechanotaxis), which is commonly observed in wound healing and cancer cell metastasis with various types of signalling molecules. In wound healing, immune cells are directed to chase a bacterium or virus; high concentration of transforming growth factor-beta (TGF-beta) induces the migration of (myo)fibroblasts towards the wound from the uninjured skin Cumming et al. (2009), Enoch and Leaper (2008). Cancer metastasis is triggered by cancer cell proliferation and cell migration. On a cellular level, chemotaxis or haptotaxis is an essential cue for cancer cell migration and hence for the dissemination of tumours Wek and Staschke (2010). Numerous studies indicate that the availability of oxygen and nutrients is one of most crucial factors for the growth of tumours Roussos et al. (2011). During tumour growth, the concentration of oxygen and nutrients depletes in the vicinity of the tumour Siemann and Horsman (2015). Therefore, cancer cells have a tendency to migrate towards regions with higher concentrations of oxygen and nutrients.

Point sources and forces are modelled by the use of Dirac delta distribution in a d–dimensional framework. Let $\Omega \subset \mathbb{R}^d$ be an open region, then this distribution is defined by the following two characteristics:

1. $\delta(\boldsymbol{x}) = 0, \quad$ for all $\boldsymbol{x} \in \mathbb{R}^d \setminus \{\boldsymbol{0}\}$;
2. $\int_\Omega \delta(x) d\Omega = 1, if 0 \in \Omega.$

The biophysical interpretation of the Dirac delta distribution is that the cell exerts force by the focal adhesion points. Since these points are many orders of magnitude smaller than the mesh size in the computational domain, we assume their sizes to be negligible. For this reason, we consider point forces by the use of Dirac delta distributions. Regarding the chemical signal, which makes the cancer cells move, we consider a point source. This is just a working hypothesis, since this could be changed to any type of source.

Together with the reaction-diffusion equation, the concentration of the signal is determined by:

$$\frac{\partial c(\boldsymbol{x},t)}{\partial t} + \nabla \cdot (\boldsymbol{v} c(\boldsymbol{x},t)) - \nabla \cdot (D \nabla c(\boldsymbol{x},t)) = k\delta(\boldsymbol{x}(t) - \boldsymbol{x}_s), \boldsymbol{x} \in \Omega, t > 0, \tag{1}$$

where $c(\boldsymbol{x},t)$ is the concentration of the signalling molecule, $D$ is the diffusion rate which has been taken constant in the current study, $k$ is the secretion rate of the signal source, $\boldsymbol{x}_s$ is the position of the source, and $\boldsymbol{v}$ is the displacement velocity of the substrate that results from the cellular forces exerted on their surroundings. The velocity is computed by solving the balance of the momentum, which will be discussed in the Sect. 2.2.

Initially, we assume there is no signalling molecules over the computational domain, that is,

$$c(\boldsymbol{x}, 0) = 0, \text{ in } \Omega, t = 0.$$

As a boundary condition, we use the following Robin condition

$$\frac{\partial c}{\partial \boldsymbol{n}} + \kappa_s c = 0, \text{ on } \partial\Omega, t > 0,$$

which deals with a balance between the diffusive flux across the boundary and the flux between the boundary and the region far away from the domain of computation. The symbol $\kappa_s$, which is non-negative, represents the mass transfer coefficient. Note that as $\kappa_s \to 0$ then the Robin condition tends to a homogeneous Neumann condition, which represents no flux (hence isolation). Whereas as $\kappa_s \to \infty$ represents the case that $c \to 0$ on the boundary, which, physically, is reminiscent to having an infinite mass flow rate at the boundary into the surroundings. The Robin condition, also referred to as a mixing boundary condition, is able to deal with both these two limits and all cases between these limits.

### 2.2 Passive convection of substrate

In wound healing, (myo)fibroblasts exert forces on their direct environment, i.e. extracellular matrix, which result into contraction of the tissue Cumming et al. (2009), Enoch and Leaper (2008), Haertel et al. (2014), Li and Wang (2011). For cancer cells, Massalha and Weihs (2016) indicate that the metastatic cells exert traction forces ranged from $100 - 600$ nN on the gel, of which the Young's modulus ranged from $2.2 - 10.9$ kPa. Furthermore, for stiffer substrates, the cancer cells remain rounded with changing area and they exert large traction forces with large magnitudes to its direct environment. Hence, the model includes passive convection of the substrate, which can provide a more realistic model in various applications.

As the cell membrane is broken into multiple line segments by nodal points, point forces are implemented here to depict the forces exerted by the cell, which are applied on the midpoint of each line segment; see Fig. 2 as an example of a square-shape cell. Among different applications, the force direction may differ. For example, if the cell encounters an obstacle, a repulsive force will be exerted to resist the compression of the cell; in wound contraction, (myo)fibroblasts exert pulling forces on the extracellular matrix (ECM).

Morphoelasticity is widely used in the biological modelling to describe elastic growth, for instance, the growth of tumours (Goriely and Moulton 2011), the seashell growth (Rudraraju et al. 2019), large contractions in wound healing (Koppenol 2017; Ben Amar et al. 2015), etc. In wound healing, morphoelasticity describes the phenomena when the deformation of the skin is so large that the deformations are plastic. Conservation of momentum, combined with the evolution equation for the effective Eulerian strain, results into the following modelling equations Koppenol (2017):

$$\begin{cases} \rho[\frac{\mathrm{D}\boldsymbol{v}}{\mathrm{D}t} + \boldsymbol{v}(\nabla \cdot \boldsymbol{v})] - \nabla \cdot \boldsymbol{\sigma} = \boldsymbol{f}, \text{ in } \Omega, t > 0, \\ \frac{\mathrm{D}\boldsymbol{\epsilon}}{\mathrm{D}t} + \boldsymbol{\epsilon}\,\mathrm{skw}\,(\boldsymbol{L}) - \mathrm{skw}\,(\boldsymbol{L})\boldsymbol{\epsilon} + [\,\mathrm{tr}\,(\boldsymbol{\epsilon}) - 1]\,\mathrm{sym}\,(\boldsymbol{L}) \\ \quad = -\alpha\boldsymbol{\epsilon}, \text{ in } \Omega, t > 0, \\ \boldsymbol{v}(\boldsymbol{x}, t) = \boldsymbol{0}, \text{ on } \partial\Omega, t > 0, \end{cases} \tag{2}$$

where $\rho$ is the density of the extracellular matrix, $\boldsymbol{L} = \nabla\boldsymbol{v}$ and $\alpha$ is a non-negative constant. Note that if $\alpha = 0$, then as soon as the force $\boldsymbol{f} = \boldsymbol{0}$, then the tissue will gradually recover to its original shape and volume. Here, $\frac{\mathrm{D}\boldsymbol{y}}{\mathrm{D}t} = \frac{\partial \boldsymbol{y}}{\partial t} + \boldsymbol{v}\nabla \cdot \boldsymbol{y}$ is material derivative where $\boldsymbol{y}$ is any tensor field and $\boldsymbol{v}$ is the migration velocity of any point within the domain of computation. In order to have a fixed boundary, we use a homogeneous Dirichlet boundary condition for the velocity. This condition implies that the overall domain boundary $\Omega$ does not deform. Hence, the overall $\Omega$ is constant over time.

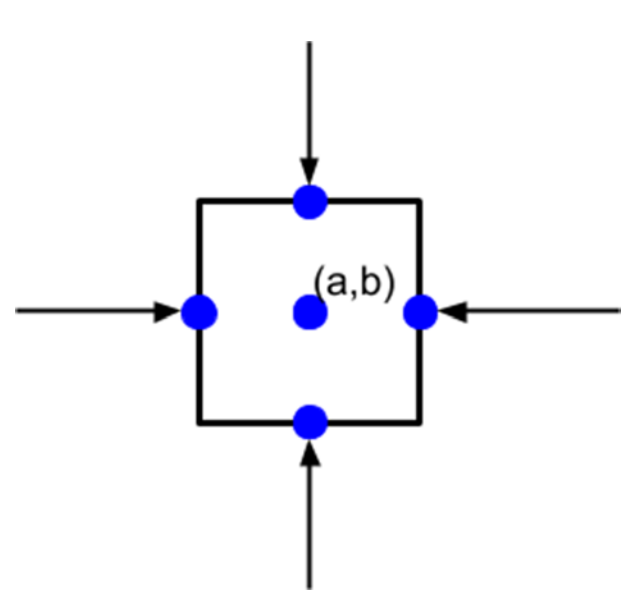

**Fig. 2** An example of pulling point forces, tractions applied by a cell, in which the cell membrane is split by four nodal points

However, due to the cellular traction forces, the tissue, and hence the computational domain, is subject to local deformations, which result into local strains and stresses. These deformations, indeed, give rise to local displacements within the domain of computation. Since the momentum balance equations are solved over the entire domain of computation, the local deformations, stresses and displacements are taken into account over the entire domain of computation. These local displacements induce the passive convection term $c(\boldsymbol{x}, t)(\nabla \cdot \boldsymbol{v})$ in Eq. (1). From a mechanical point of view, we treat the computational domain as a continuous linear isotropic domain. Further, as a result of the presence of liquid phases in the tissue, the mechanical balance is also subject to viscous, that is friction, effects. Therefore, we use Kelvin–Voigt's viscoelastic dashpot model, of which the stress tensor reads as

$$\begin{aligned} \boldsymbol{\sigma} &= \boldsymbol{\sigma}_{elas} + \boldsymbol{\sigma}_{visco} \\ &= \frac{E}{1 + \nu_s}\left\{\boldsymbol{\epsilon} + \mathrm{tr}\,(\boldsymbol{\epsilon})\left[\frac{\nu_s}{1 - 2\nu_s}\right]\boldsymbol{I}\right\} \\ &\quad + \mu_1\,\mathrm{sym}\,(\boldsymbol{L}) + \mu_2\,\mathrm{tr}\,(\,\mathrm{sym}\,(\boldsymbol{L}))\boldsymbol{I}, \end{aligned} \tag{3}$$

where $\nu_s$ is the Poisson's ratio of the substrate, $\boldsymbol{\epsilon}$ is the strain tensor, $\mu_1$ and $\mu_2$ are the shear and bulk viscosity, respectively. The morphoelasticity model solves nonlinear equation and both velocity $\boldsymbol{v}$ and strain tensor $\boldsymbol{\epsilon}$ are unknowns. The deformation of the domain is actually determined by the strain tensor. The displacement of the domain can be approximated by integrating the velocity over time: $\boldsymbol{u}(t) \approx \int_0^t \boldsymbol{v}(s)\mathrm{d}s$, where the velocity is determined by Eq. (2). The approximated solution is then known on the moving finite-element meshpoints. This is the discrete counterpart of the displacement. Further, we need the displacement (velocity) at the positions of the nodal points on the cell boundary. Since these positions do not coincide with the positions of the finite-element meshpoints, we need a mapping from the displacement velocity obtained at the finite-element meshpoints onto the, continuous, positions of the cell boundary nodes. This is obtained through interpolation procedures based on the Lagrangian finite-element framework. The obtained displacement velocities are substituted into the equation for the migration of the cell boundary nodes.

In the application of wound healing, (myo)fibroblasts are the cells pulling the ECM and causing the contractions. The traction force $\boldsymbol{f}$ of each (myo)fibroblast reads as

$$\mathbf{f}(\boldsymbol{x};t) = \sum_{j=1}^{N} P(\boldsymbol{x}_j;t)\boldsymbol{n}(\boldsymbol{x}_j(t))\delta(\boldsymbol{x}(t) - \boldsymbol{x}_j(t))\Delta\Gamma^j, \quad \boldsymbol{x} \in \Omega, t0, \tag{4}$$

where $N$ is the number of nodal points on the cell membrane, $P(\boldsymbol{x};t)$ is the magnitude of the force exerted by each (myo)fibroblast per unit length of the cell membrane. We have constructed the model such that local differences of the cellular traction force over the cell boundary can be incorporated. In the current study, however, we have assumed the cellular traction force to be constant over the entire cell boundary and for all the cells. Furthermore, $\boldsymbol{n}(\boldsymbol{x}(t))$ is the unit inward pointing normal vector (towards the cell centre) at $\boldsymbol{x}(t)$ (see Fig. 2 as a schematic), $\boldsymbol{x}_j(t)$ is the midpoint of line segment $j$, and $\Delta\Gamma^j$ is the length of line segment $j$. Note that here $\boldsymbol{x}$ refers to any position in $\Omega$, hence possibly, but not necessarily on the cell boundary, or on the finite-element mesh points. The positions $\boldsymbol{x}_j(t)$ represent the positions of the nodal points on the cell boundary (hence not on the finite-element mesh), where the cellular traction force is exerted.

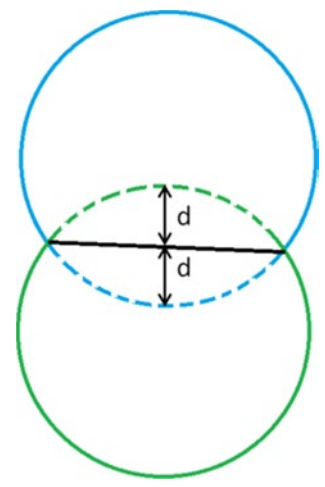

**Fig. 3** When cells collide with each other, they will deform and exert repulsive forces. The dashed curves show the equilibrium shape of cells, and the black curve is the overlapping membrane of both cells

Further, we consider (myo)fibroblasts colliding with each other, then the cells exert repelling forces on the other one. Here, cells are not allowed to intersect each other; see Fig. 3 as a schematic. Suppose for (myo)fibroblast $i$, there are $N_m^i = \{j_1^i, \dots, j_m^i\}$ line segments of cell membrane mechanically contacting with other (myo)fibroblasts. Then, online segments $j \in N_m$, the (myo)fibro-blast exert repelling force, while on the rest of the line segments, (myo)fibroblast releases pulling forces on the ECM. Hence, the traction force of the (myo) fibroblast $i$ is given by

$$\begin{aligned}\mathbf{f}^i(\boldsymbol{x};t) = &\sum_{j=1, j\notin N_m^i}^{N} P(\boldsymbol{x}_j^i, t)\boldsymbol{n}(\boldsymbol{x}_j^i(t))\delta(\boldsymbol{x}(t) - \boldsymbol{x}_j^i(t))\Delta\Gamma^{i,j} \\ &- \sum_{j\in N_m^i} Q(\mathrm{d}(\boldsymbol{x}_j^i), t)\boldsymbol{n}(\boldsymbol{x}_j^i(t))\delta(\boldsymbol{x}(t) - \boldsymbol{x}_j^i(t))\Delta\Gamma^{i,j},\end{aligned} \tag{5}$$

where $l_m$ is the portion of the (myo)fibroblast membrane mechanically contacting with other cell, $Q(\mathrm{d}(\boldsymbol{x}), t)$ is the force magnitude per length, and $\mathrm{d}(\boldsymbol{x})$ is the penetration depth. According to two-dimensional Hertz theory Popov (2010), Liu et al. (2005), Tripp (1985), for each elastic body, the explicit relation between the total force and the penetration depth is not clear. We assume the total force magnitude $\tilde{Q}(\mathrm{d}(\boldsymbol{x}), t)$ is linearly proportional to the penetration depth $\mathrm{d}(\boldsymbol{x})$:

$$\tilde{Q}(\mathrm{d}(\boldsymbol{x}), t) = \frac{\pi}{4}\mathrm{d}(\boldsymbol{x})E^*, \tag{6}$$

where $E^*$ is the total equivalent Young's modulus derived by

$$\frac{1}{E^*} = \frac{1-\nu_1^2}{E_1} + \frac{1-\nu_2^2}{E_2}.$$

Here, $\nu_i$ and $E_i$ with $i \in \{1, 2\}$ represent the Poisson ratio and Young's modulus of two elastic bodies, respectively. In particular, if two bodies have the same elastic characteristics (i.e. $\nu_1 = \nu_2 = \nu$ and $E_1 = E_2 = E$), then

$$E^* = \frac{E}{2(1-\nu^2)}.$$

We assume that the magnitudes of the repulsive force, which is exerted on the boundary segments of cell $i$ that are in contact with another cell, are identical. In other words, $Q(\mathrm{d}(\boldsymbol{x}), t)$ is given by

$$Q(d(\boldsymbol{x}), t) = \frac{\tilde{Q}(\mathrm{d}(\boldsymbol{x}), t)}{||l_m||} = \frac{\pi}{4}\mathrm{d}(\boldsymbol{x})E^*/||l_m||, \tag{7}$$

where $||l_m||$ is the total length of the portion of the membrane of (myo)fibroblast $i$ mechanically contacting with other (myo)fibroblast (i.e. the sum of the length of $\Delta\Gamma^{i,j}, j \in N_m^i$). Subsequently, the total traction force is $\boldsymbol{f} = \sum_{i=1}^{N_C} \boldsymbol{f}^i$, where $N_C$ is the number of (myo)fibroblasts that are in contact with each other.

Another application is metastasis and invasion of cancer cell. Usually, in vitro, a microtube experiment is conducted Mak et al. (2013). To simplify the model, here we only consider one cancer cell going through a microtube; see Fig. 4. Similar to the case when (myo)fibroblasts collide, we assume that $N_m = j_1, \dots, j_m$ is the line segments of cell membrane mechanically contacting with the wall of the microtube. Here, we will exclude the pulling force. In other words, the force released by the cancer cell is only the repelling force exerted on the wall of the microtube:

$$\mathbf{f}_m(\boldsymbol{x};t) = -\sum_{j\in N_m} Q_m(\mathrm{d}(\boldsymbol{x}_j), t)\boldsymbol{n}(\boldsymbol{x}_j(t))\delta(\boldsymbol{x}(t) - \boldsymbol{x}_j(t))\Delta\Gamma^j. \tag{8}$$

The magnitude of the force $Q_m(\mathrm{d}(\boldsymbol{x}), t)$ here follows the same definition as in Eq. (7), where $\mathrm{d}(\boldsymbol{x})$ is the radius subtracting the distance from the cell membrane to the cell centre.

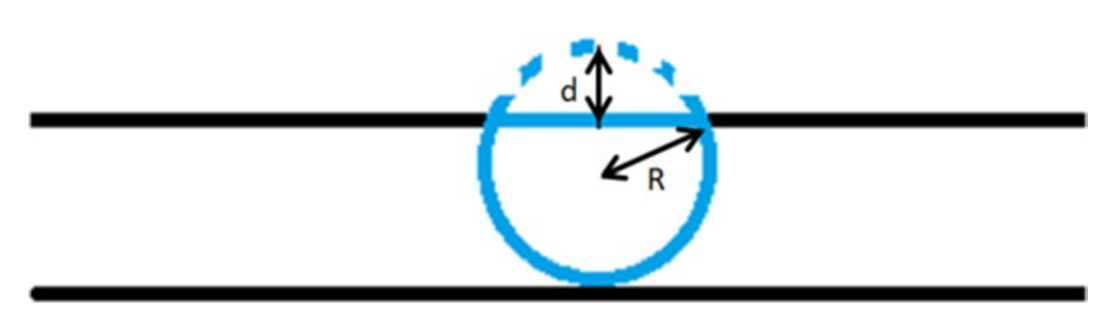

**Fig. 4** The cell is compressed by the wall of the microtube. To inhibit any changes from its equilibrium shape, it exerted the repelling force on the wall of the microtube, which is proportion to the compressed distance $d$

### 2.3 Cell deformation

According to Chen et al. (2018), Vermolen and Gefen (2012), the cell cytoskeleton is depicted as a collection of springs between the centre position of the cell and the nodal points on the cell membrane. Therefore, the equilibrium shape of the cell is kept by these springs, regardless of the original cell shape. In this manuscript, we consider circular, elliptic and star-shape cells as two-dimensional projections. Combining chemotaxis (or mechanotaxis), passive convection and random walk, the displacement of the nodal point $j$ on the cell membrane is given by

$$d\boldsymbol{x}_j = \beta(CSI(\Omega_C), A(\Omega_C))\frac{\nabla c(\boldsymbol{x}_j, t)}{\|\nabla c(\boldsymbol{x}_j, t)\| + \gamma}\mathrm{d}t \\ + E_c(\boldsymbol{x}_c(t) + \hat{\boldsymbol{x}}_j - \boldsymbol{x}_j(t))\mathrm{d}t + \boldsymbol{v}\mathrm{d}t + \sigma_{rw}\mathrm{d}\boldsymbol{W}(t), \\ \text{in } \Omega_m \subset \Omega. \tag{9}$$

Here, $\Omega_C$ represents the cell region, $\Omega_m$ is the domain occupied by the microtube, $E_c$ represents the cell elasticity; $\hat{\boldsymbol{x}}_j = \tilde{\boldsymbol{x}}_j(t) - \boldsymbol{x}_c(t)$ is the vector connecting the equilibrium position of nodal point $i$ on the cell membrane to the cell centre, $\boldsymbol{x}_c$ is the central position of the cell, and $\tilde{\boldsymbol{x}}_j$ represents the equilibrium position of the nodal point $j$ corresponding to the cell centre $\boldsymbol{x}_c$ (see Fig. 1); $\gamma$ is a small positive constant to prevent the denominator being zero; $\boldsymbol{v}$ is the velocity of the substrate determined by Eq. (2); $\sigma_{rw}$ is the portion of random walk, and $\mathrm{d}\boldsymbol{W}(t)$ is a vector-Wiener process, which accounts for random walk. Furthermore, $\beta = \beta(CSI(\Omega_C), A(\Omega_C))$ is the weight of chemotaxis (or mechanotaxis), where we define the Cell Shape Index (CSI) of cell $\Omega_C$ by

$$CSI(\Omega_C) = \frac{4\pi A(\Omega_C)}{l^2(\partial\Omega_C)},$$

where $A(\Omega_C)$ is the cell area, $l(\partial\Omega_C)$ is the circumference of the cell membrane. According to Keren et al. (2008), reduction of cell area and deformation of cell shape reduce the mobility of cell. For simplicity, we propose a linear relation here:

$$\beta(CSI(\Omega_C), A(\Omega_C) = \beta_0 \times \mu_m \\ \times (CSI(\Omega_C)/CSI_0(\Omega_C) + A(\Omega_C)/A_0(\Omega_C))/2,$$

where $\beta_0$ is the maximal response from the cell to the signal, $\mu_m$ is the mobility reduction coefficient, and $CSI_0(\Omega_C)$ and $A_0(\Omega_C)$ represent the CSI and volume of the equilibrium cell.

In order to maintain the right orientation of the cell, we introduce a matrix after rotation of an angle $\phi$, as in Chen et al. (2018):

$$\boldsymbol{B}(\phi) = \begin{pmatrix} \cos(\phi) & -\sin(\phi) \\ \sin(\phi) & \cos(\phi) \end{pmatrix}, \tag{10}$$

such that $\phi$ can be computed from

$$\tilde{\phi} = \arg \min_{\phi \in [0, 2\pi)} \left( \sum_{i=1}^{N} \|\boldsymbol{B}(\phi)\tilde{\boldsymbol{x}}_i(t) - \boldsymbol{x}_i(t)\|^2 \right). \tag{11}$$

The orientation of the cell is important in the context of how the cell has rotated from its initial position. The orientation of the cell is represented by the angle of the vector connecting the “front and tail” of the cell. The overall displacement of the nodes of the cell boundary is determined by translation and rotation. This matrix $\boldsymbol{B}(\phi)$ monitors the angle of rotation of the cell with respect to the cell position (and hence boundary nodes) at the previous time step. This orientation and hence the angle of rotation is important for the determination of the equilibrium points of the cell boundary nodes. The equilibrium points reflect the position to which the cell boundary nodes will converge to if the cell does not move (that is the chemical signal is set to zero), and if the extracellular matrix is not subject to displacement velocities. If this orientation, that is the angle, would not be incorporated, then the cell with always return to its initial orientation.

Hence, the displacement of nodal point $j$ is adapted to

$$\mathrm{d}\boldsymbol{x}_j = \beta(CSI(\Omega_C), A(\Omega_C))\frac{\nabla c(\boldsymbol{x}, t)}{\|\nabla c(\boldsymbol{x}, t)\| + \gamma}\mathrm{d}t \\ + E_c(\boldsymbol{x}_c(t) + \boldsymbol{B}(\tilde{\phi})\hat{\boldsymbol{x}}_j - \boldsymbol{x}_j(t))\mathrm{d}t + \boldsymbol{v}\mathrm{d}t + \sigma_{rw}\mathrm{d}\boldsymbol{W}(t). \tag{12}$$

If there is an obstacle encountered by the cell, adjusting the displacement is necessary. Denote $\partial\Omega_{ob}$ as the boundary of the obstacle, which is possibly another cell or the wall of the microtube. For the nodal point colliding the obstacle, it cannot pass over the boundary of the obstacle. Hence, for the displacement of nodal point $j$, the normal direction of the boundary of the obstacle in $\mathrm{d}\boldsymbol{x}_j(t)$ is vanished. To rephrase it, we adjust the displacement of the nodal point if it collides the obstacle (see Fig. 5) by

$$d\boldsymbol{x}_j(t) \leftarrow d\boldsymbol{x}_j(t) - (\mathrm{d}\boldsymbol{x}_j(t), \boldsymbol{n}_{ob}(\boldsymbol{x}_j(t)))\boldsymbol{n}_{ob}(\boldsymbol{x}_j(t)), \quad \text{if } \boldsymbol{x}_j(t) \in \partial\Omega_{ob}, \tag{13}$$

where $\boldsymbol{n}_{ob}(\boldsymbol{x})$ is unit pointing normal vector (outward the cell centre).

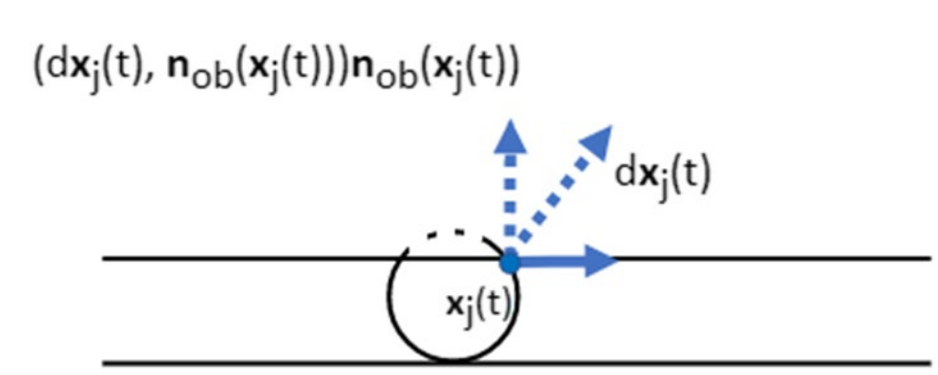

**Fig. 5** A schematic to show that the adjustment of the displacement of the nodal point on the cell membrane, when the nodal point contacts the wall of the microtube

Furthermore, if a cell tries to go through a microtube, which mimic the cancer cell metastasis and invasion, not only the direction of the cell migration is limited (since the cell cannot pass over the microtube), but also the microtube will slow down the velocity of the cell as a result of friction. Note that the magnitude of the friction is proportional to the repelling force. In our model, we simply distract part of the velocity in the tangential direction of the obstacle. Hence, the displacement of the nodal point which collides the wall of the microtube is given by

$$\mathrm{d}\boldsymbol{x}_j(t) \leftarrow \mathrm{d}\boldsymbol{x}_j(t) - \mu_f \|\boldsymbol{f}(\boldsymbol{x}_j(t))\| \times (\mathrm{d}\boldsymbol{x}_j(t), \boldsymbol{\tau}_{ob}(\boldsymbol{x}_j(t)))\boldsymbol{\tau}_{ob}(\boldsymbol{x}_j(t)), \ \text{if } \boldsymbol{x}_j(t) \in \partial\Omega_{ob}, \tag{14}$$

where $\mu_f$ is the cell friction coefficient, $\mathbf{f}(\boldsymbol{x}_j(t))$ is the repelling force exerted by the cell and $\boldsymbol{\tau}_{ob}(\boldsymbol{x})$ is the tangential direction of the obstacle boundary $\partial\Omega_{ob}$.

This model provides a simple computational framework to describe the dynamics of the cell shape under multiple circumstances. However, the model does not describe the Poisson effect of the cell if the cell is compressed, since the model mainly considers the extracellular environment. Hence, in this manuscript, the cell length will not be investigated.

## 3 Applications and numerical results

We exhibit several possible applications in this section, namely, cells migrating as a result of chemotactic signals, cells differentiating to another phenotype, cells repelling each other and one cell migrating through a microtube. Some parameters are the same in every application. If there is no specification, the parameter values are shown in Table 1. Note that parameter values are partially determined by experimental data from the references and partially estimated in this study, as they are indicated in all the parameter tables. We try to use the clinical/experimental data from the literature as much as possible; however, some parameter values are unknown. Hence, to estimate theses unknown parameter values, we determined the value by reproducing the experiment as much as possible.

In particular, to validate and calibrate the model, we tried to reproduce the key results in Mak et al. (2013) like the probability of the occurrence of Phase 3 and the time interval of each phase. We ran four different Monte Carlo simulations to calibrate the model and to see the impact of different settings of the model.

### 3.1 Finite-element methods

In this manuscript, all the boundary value problems are solved by the finite-element methods with Lagrange linear basis functions. Regarding the time-integration, we use a backward Euler method. From the theory, it is known that smooth solutions would be subject to errors of the order $\mathcal{O}(h^2)$ in the $L^2$–norm of the numerical approximation and $\mathcal{O}(\Delta t)$.

**Table 1** Parameter values used in all the applications

| Parameter | Description | Value | Units | Source |
|---|---|---|---|---|
| $E_s$ | Substrate elasticity | 100 | $\mathrm{kg}/(\mu\mathrm{m}\cdot\mathrm{min}^2)$ | Liang et al. (2010) |
| $E_c$ | Cell elasticity | 5 | $\mathrm{kg}/(\mu\mathrm{m}\cdot\mathrm{min}^2)$ | Chen et al. (2017) |
| $\mu_f$ | Cell friction coefficient | 0.03 | – | Angelini et al. (2012) |
| $\nu_s$ | Poisson's ratio of the ECM | 0.49 | – | Koppenol (2017) |
| $\nu_c$ | Poisson's ratio of (myo)fibroblast and cancer cell | 0.32 | – | Trickey et al. (2006) |
| $k$ | Secrete rate of the signal | 2.5 | $\mathrm{kg}/(\mu\mathrm{m}^3\cdot\mathrm{min})$ | Peng and Vermolen (2020) |
| $\kappa_s$ | Parameter in Robin's boundary condition to solve Eq. (1) | 100 | – | Peng and Vermolen (2020) |
| $\mu_1$ | Shear viscosity of the ECM | 33.783 | – | Peng and Vermolen (2020) |
| $\mu_2$ | Bulk viscosity of the ECM | 22.523 | – | Peng and Vermolen (2020) |
| $\beta_0$ | Maximal mobility of points on cell membrane | 10 | $\mathrm{min}^{-1}$ | Estimated in this study |
| $N$ | Number of nodal points on the cell membrane | 40 | – | Estimated in this study |
| $\mu_m$ | The coefficient of cell mobility reduction | 1 | – | Estimated in this study |
| $\sigma_{rw}$ | Weight of random walk | 1 | – | Estimated in this study |
| $\alpha$ | Degree of permanent deformation in Eq. (2) | 0.1 | – | Estimated in this study |

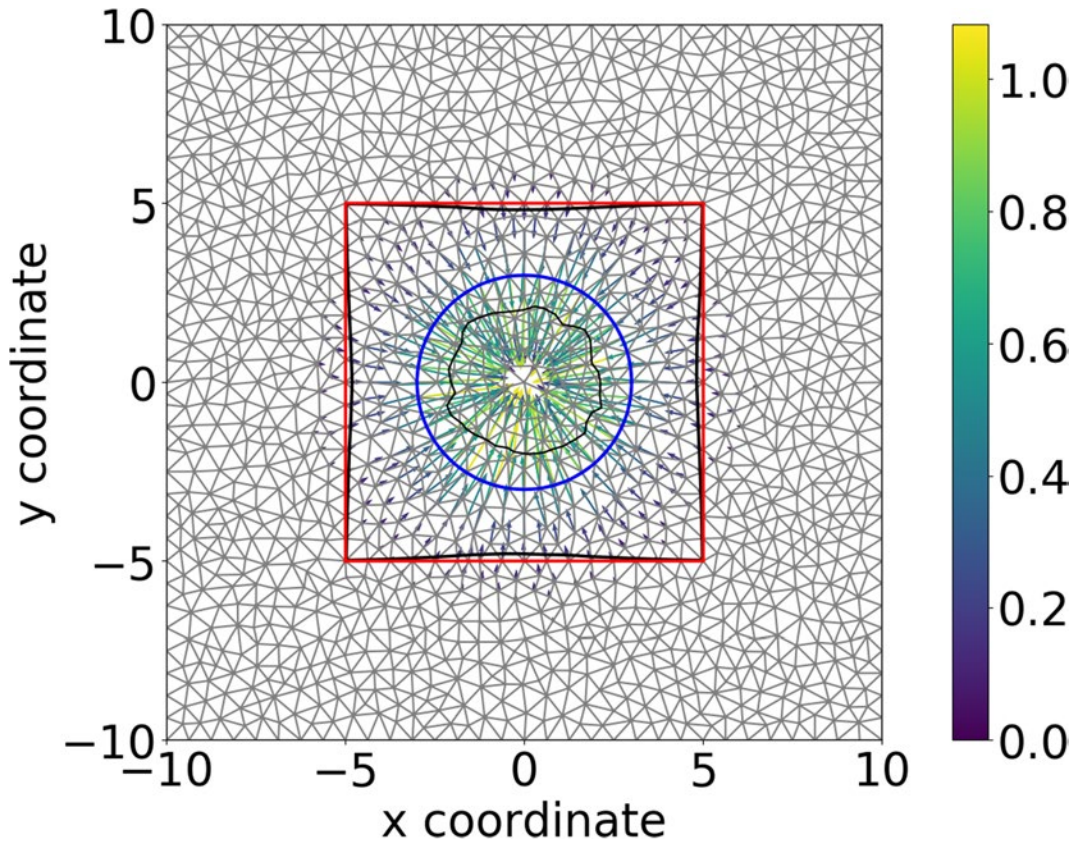

**Fig. 6** The plot shows the solution Eq. (2), as an example to investigate the convergence of the finite-element methods. Blue curve represents the initial cell membrane, and red curve represents the original subdomain boundary. Black curves show the deformed shape of the cell and the subdomain

Here, we solved the boundary value problem in Eq. (2) with the traction forces expressed in Eq. (4). For the sake of investigating the convergence of the finite-element method, the parameters are dimensionless. We consider one large non-moving cell in the computational domain (see Fig. 6), of which the membrane is divided into finite line segments, and there is a traction force applied on the midpoint of each line segment as in Eq. (4). With the refinement of the mesh, the convergence rate of $L^2$–norm of the solution to Eq. (2) (i.e. the velocity) is computed, which is 1.899828112 that is close to the theoretical value 2.

## 3.2 Cells moving towards the point source

The basic application is that cell migrates towards the concentration gradient of a signalling molecule, which can be oxygen, growth factors or virus. The displacement is mainly determined by the gradient of the concentration of the signal. Subsequently, the closest part of the cell to the emitting source will develop a "nose"; see Fig. 7. All parameter values are documented in Tables 1 and 2. The "nose" behaviour (or the so-called a triangular tailed shape) has also been observed in biological experiments, which is due to the locations of the adhesion sites over the cell membrane Mogilner and Keren (2009). In the simulation, we accommodate for the engulfment of the chemical source by switching off the chemical signal once the cell physically contacts it. Once this signal has been switched off, the concentration gradient flattens as a result of diffusion processes and therewith the cell recesses back to its equilibrium (original) shape and volume. At that moment, the cell geometry is no longer determined by chemotaxis.

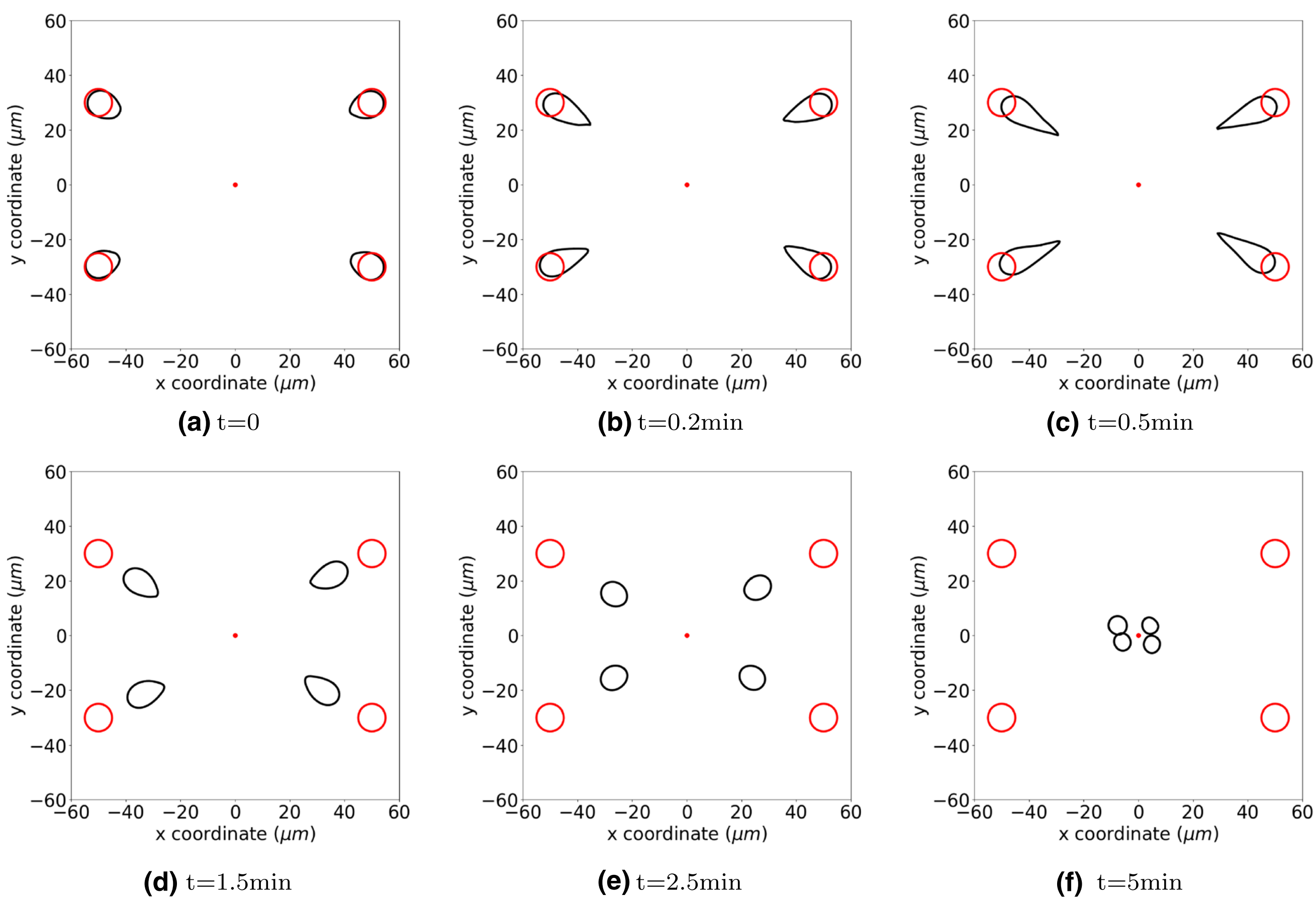

**Fig. 7** The figure shows the positions of cells at consecutive times. The red circles show the original shape and position of cells, and the red dots in the centre is the point source of a generic signal

From $t = 0$, the cells are attracted towards the centre of the computational domain, which is the location of the source of signalling molecules. Due to the difference of the gradient of the concentration of the signal over the domain, cells are deformed into a droplet shape, where the "nose" points in the direction of the point source. As the diffusion of the signal proceeds, the "nose" gradually disappears and cells recover to the equilibrium shape. To evaluate the cell geometry quantitatively, we provide the evolution of the CSI and cell area as a function of time in Fig. 8. These quantities are of interest from a clinical point of view Keren et al. (2008). Resulting from the displacement of the direct environment, the volume of the cell decreases. The permanent volume changes of the cell are imposed by the permanent displacements from the morphoelastic model. Furthermore, the cells are tethered within a rigid deformed structure; hence, it makes cells deform as well. We note that cells have already recovered to the original shape but not the volume. It can be implied that a stiffer cell deforms less compared to a softer one.

### 3.3 Differentiation of cells

Cell differentiation is a process of a cell changing from one phenotype to another one, for example, a stem cell differentiates into various phenotypes, like blood cells and nerve cells, etc. In this manuscript, we mainly focus on the cellular differentiation in wound healing. In the proliferative phase of wound healing, some regular fibroblasts (which are spindle-shaped Chaudhari 2015; Phan 2008) differentiate into myofibroblasts (which are dendritic-shaped Hirahara et al. 2004; Desai et al. 2014), which pull the skin ever harder and cause the permanent contractions. It is commonly recognized that high concentration of TGF-beta induces the fibroblast-to-myofibroblast differentiation Keren et al. (2008), Cumming et al. (2009), Desai et al. (2014). In this section, since we mainly want to present a model of differentiation, only differentiation from fibroblasts to myofibroblasts is exhibited as an example: the signal is TGF-beta, and initially, there are four regular fibroblasts in unwounded region, which are simulated by ellipses.

We assume that the two phenotypes of cells have the same volume when they are in the equilibrium status. Here, for ellipse and hypocycloid, there are two parameters to determine each shape: long (denoted by $a_e$) and short axis's (denoted by $b_e$) determine the ellipse; the radius of basis circle (de noted by $a_h$) and rotating circle (denoted by $b_h$) determine the hypocycloid. Note that the hypocycloid-shaped cell may not be realistic, and it is mainly to show that the model is capable to model the differentiation of cells, in which mostly the cellular skeleton and geometry are altered. To develop a smooth differentiation process, we introduce a function such that each parameter changes over time:

$$\begin{cases} R_a(\omega) = a_h\omega + a_e(1-\omega), \\ R_b(\omega) = b_h\omega + b_e(1-\omega), \end{cases} \tag{15}$$

where $R_a(\omega)$ and $R_b(\omega)$ represent two parameters to determine the shape, and $\omega = 1 - \exp\{-\lambda_\omega(t - t_\omega)\}$. Here, $\lambda_\omega$ is the parameter of the exponential distribution and $t_\omega$ is the time point when the fibroblast starts differentiating.

Figure 9 presents the cells positions at different time. In this manuscript, we only consider a phenomenological modelling formulation in the sense that the cell differentiates if it is exposed to concentrations of signalling molecule that exceed a given threshold. The shape evolution is determined by the parameters in Eq. (15).

As regular fibroblasts approach to the region with high concentration of TGF-beta, some of them start differentiating into myofibroblasts gradually. Subsequently, they exert larger forces on the ECM and contractions are developed in the wound, which is marked with red curves as a subdomain. The parameter values of the simulations are from Tables 1 and 3.

### 3.4 Repulsion between two colliding cells

Cells will deform when they encounter each other or an obstacle. On the contacting surface, cells will exert a repelling force (as it is shown in Eq. (7)) to recover to its equilibrium shape. We consider two cells colliding with each other and adjust the displacement of nodal points on the cell membrane by Eq. (13). Here, cells are not allowed to intersect with each other. Hence, initially, cells are located with a small distance between each other. In Fig. 10, we present the cell positions at different times, and cells deform due to mechanical contact (hard impingement). The parameter values are from Tables 1 and 2.

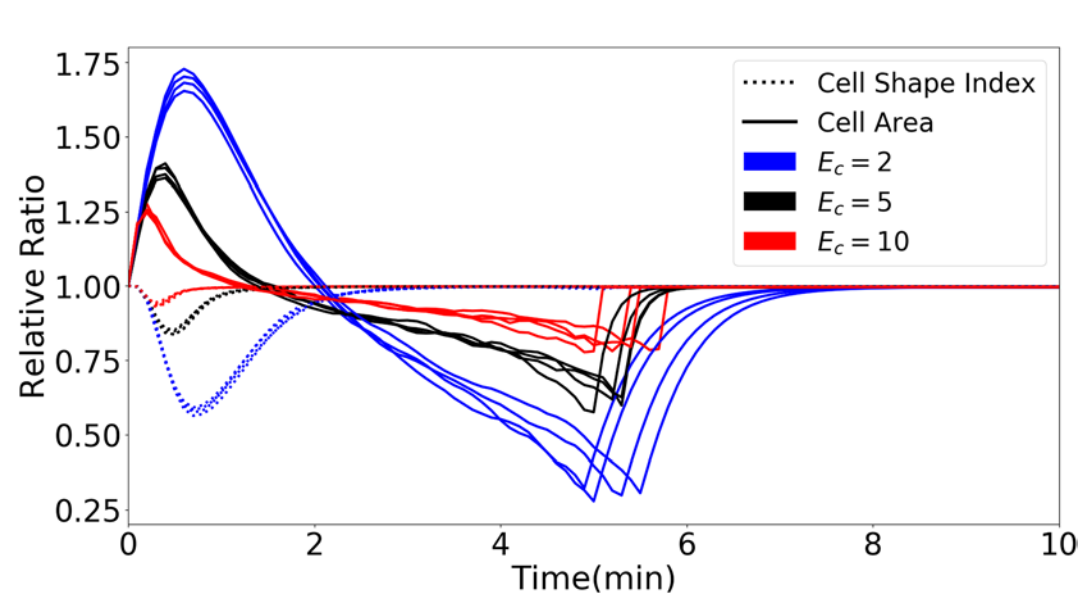

**Fig. 8** The cell shape index and relative ratio of cell area of all cells in Fig. 7 are shown in the plot. The solid curves represent the cell area, and the dashed curves are the cell shape index. Different colours of curves indicate different stiffness of the cells

**Table 2** Parameter values estimated in the application of cell migrating toward the signal source

| Parameter | Description | Value | Units |
|---|---|---|---|
| $R$ | Cell radius | 5 | μm |
| $\Delta t$ | Time step | 0.1 | min |
| $D$ | Diffusion rate of the signal | 200 | $\mu m^2/min$ |
| $x_0$ | Length of computational domain in x-coordinate | 120 | μm |
| $y_0$ | Length of computational domain in y-coordinate | 120 | μm |
| $x_w$ | Length of the subdomain in the centre of computational domain in x-coordinate | 40 | μm |
| $y_w$ | Length of the subdomain in the centre of computational domain in y-coordinate | 40 | μm |

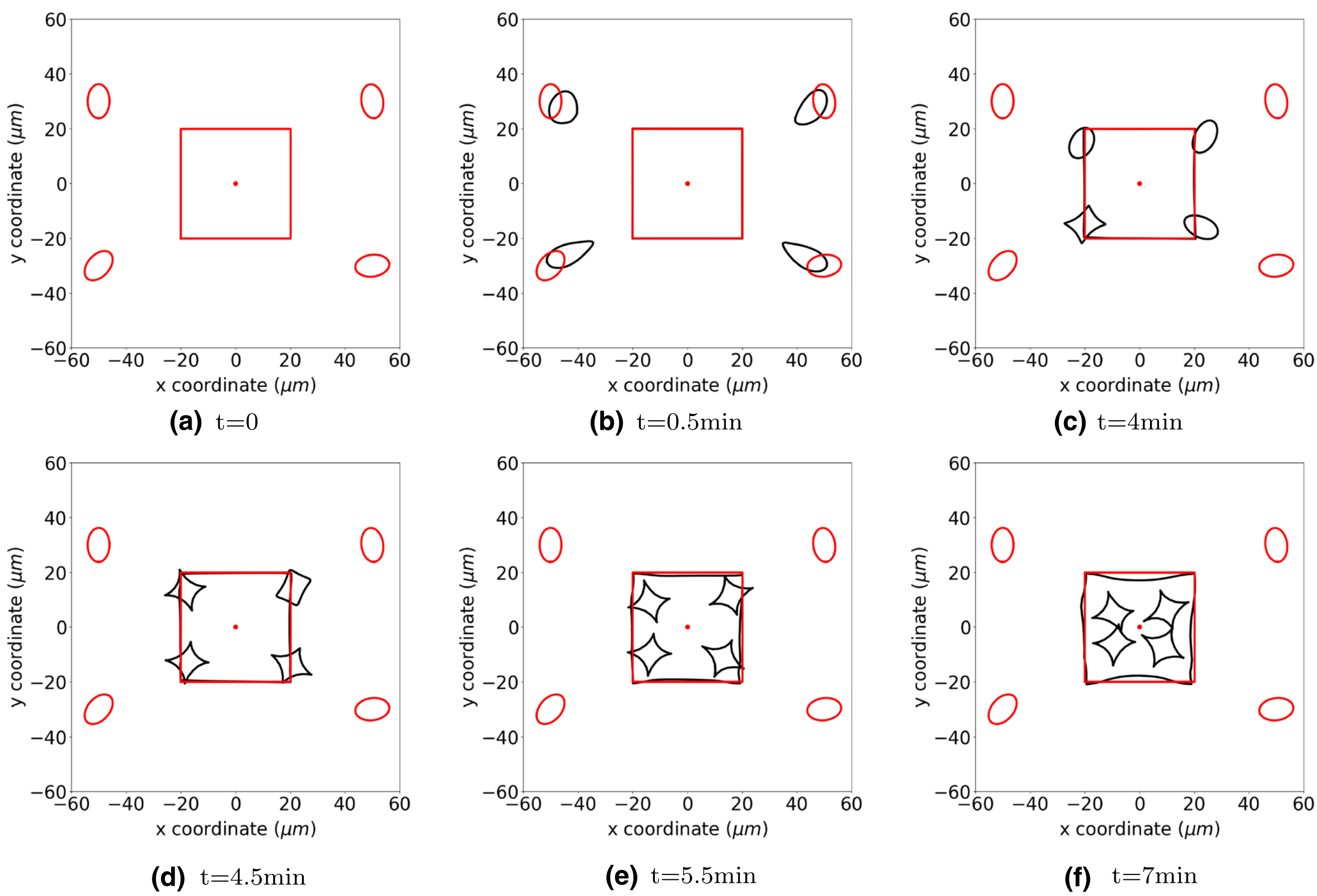

**Fig. 9** The figure shows the positions of cells at consecutive times by the black curves. The red circles show the original shape and position of cells, and the red dots in the centre is the point source of TGF-beta which triggers the differentiation from regular fibroblasts to myofibroblasts

### 3.5 Cell moving through a microtube

Metastasis is a difficult phenomena to study due to its large variation in spatiotemporal scales. Hence, studying the mechanics of one single cell is essential since the individual cell needs to break out from the tumour and invade though the ECM. To achieve that, Mak et al. (2013) developed an active microfluidic system with several features to mimic the metastasis and invasion of the cancer cell. In this section, we will use our model to reproduce results of the experiment in Mak et al. (2013).

**Table 3** Parameter values estimated in the application of cell differentiation

| Parameter | Description | Value | Units |
|---|---|---|---|
| $a_e$ | Length of long axis in elliptic cell | 6.25 | μm |
| $b_e$ | Length of short axis in elliptic cell | 4 | μm |
| $a_h$ | Radius of the basis circle to draw hypocycloid-shape cell | $20/\sqrt{6}$ | μm |
| $b_h$ | Radius of the rotating circle to draw hypocycloid-shape cell | $5/\sqrt{6}$ | μm |
| $\lambda_\omega$ | Parameter in the exponential distribution to compute $\omega$ | 10 | – |
| $\Delta t$ | Time step | 0.1 | min |
| $D$ | Diffusion rate of the signal | 233.2 | $\mu m^2/min$ |
| $x_0$ | Length of computational domain in x-coordinate | 120 | μm |
| $y_0$ | Length of computational domain in y-coordinate | 120 | μm |
| $x_w$ | Length of the subdomain in the centre of computational domain in x-coordinate | 40 | μm |
| $y_w$ | Length of the subdomain in the centre of computational domain in y-coordinate | 40 | μm |

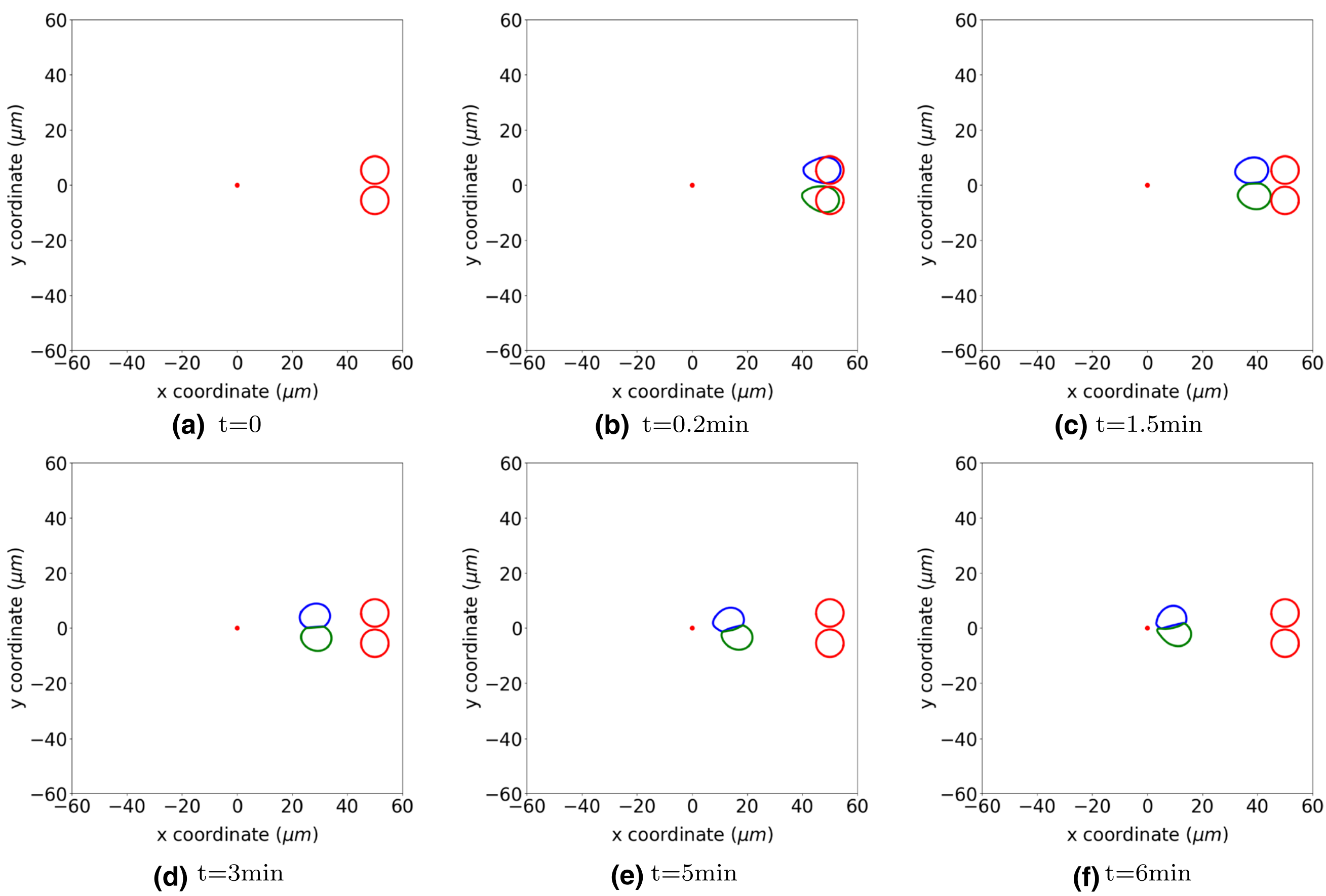

**Fig. 10** The figure shows the positions of cells in blue and green at consecutive times when two cells collide. The red circles show the original shape and position of cells, and the red dots in the centre is the point source of a signal

Mak et al. (2013) introduced a device and procedure to investigate mechanical transition effects of invading cancer cells. They constructed a set up with periodic microfluidic channels with various lengths and widths. This set up consists of a repeated pattern of large channel (LC) and subnuclear barriers (SNB). Mak et al. (2013) let cancer cells load at the reservoir. Subsequently, cells move simultaneously through the microtubes and data regarding the velocity and cell length are collected.

### 3.5.1 Simulation settings

Following the settings in the experiment, we define a microtube with a varying width: a 15 µm larger channel (LC) and 3.3 µm subnucleus barriers with length 10 µm (SNB10). Since the main reason for the active migration of the cell is not evident in Mak et al. (2013), we keep on using random walk, with either chemotaxis or fixed velocity of the cell. Rather than having a periodic setting of subnucleus barriers (SNBs), we have one SNB in the middle connected with two LCs and run the simulations, respectively. In this manuscript, we only run the simulations regarding SNB10 in Mak et al. (2013).

We consider the reduction of the cell mobility caused by cell shape and cell area Keren et al. (2008), which is explained in Eq. (12), the repelling forces exerted by the cell on the obstacles in Eq. (13) and the friction between the cell and the wall of the microtube in Eq. (14).

The position and shape of the cell are shown in Fig. 11, which indicates how the cell migrates through the microtube. Since the repelling force on the wall of the microtube is included, we investigate the results regarding the cell velocity, pressure and the cell shape index over time; see Fig. 12. The parameter values are taken from Tables 1 and 4.

Initially, there is a short distance before the cell enters the microtube; here, the cell encounters no distraction. Therefore, the cell travels at maximal speed and the cell is not compressed in the beginning. Next to this, the gradient of the signal results into the "nose" behaviour, and hence, the cell shape index changed. As the cell enters the wider part of the microtube, it slows down due to the friction, and the cell is compressed; therefore, the cell starts exerting pushing forces on the wall of the tube. In the LC part, the cell shape index stays stable around 0.95. Further, the cell approaches the SNB, which is much more narrow than the LC; the cell suffered more from the friction and the compression from the microtube. As a consequence, the minimal cell velocity, the cell shape index and the maximal pressure are recorded when the cell is in the SNB. After that, cell moves further towards the signal source through the LC again. Hence, the cell velocity and cell shape index increase again, while the cell pressure reduces. According to Angelini et al. (2012) and McCann et al. (2010), we manage to keep the cell velocity and pressure in a reasonable range: $6 - 20$ µm/min and the maximal pressure that a cell can handle is around 12 kPa.

### 3.5.2 Monte Carlo simulations

In Mak et al. (2013), the displacement of the cell is categorized as four phases:

- *Phase 1* The cell enters the microtube via the LC and slows down in particular when it is approaching the SNB;
- *Phase 2* The cell is compressed strongly to enter the SNB;
- *Phase 3* The cell fails to migrate monotonically forward when it is in the SNB;
- *Phase 4* The cell enters the LC again and continues to migrate monotonically.

Hence, in the simulations, we try to collect the data and distinguish these different phases. Different from Mak et al. (2013) that the microtube is designed periodically (such that the sample can be collected multiple times with one individual cell), one cell is supposed to go through one set of the microtube in each simulation. To reproduce the experimental results, Monte Carlo simulations are conducted to estimate the probability of the occurrence of phase 3 and the time cost for each phase, with different aforementioned reasons of active migration. The input values for the Monte Carlo simulations are shown in Table 5. In our simulation, we determine phase 3 when the cell stops moving monotonically forward when it is in the SNB until it leaves the SNB completely and reenters the LC.

**Table 4** Parameter values used in the application of cell going through a microtube

| Parameter | Description | Value | Units | Source |
|---|---|---|---|---|
| $R$ | Cell radius | 9 | µm | Mak et al. (2013) |
| $\Delta t$ | Time step | 0.07 | min | Estimated in this study |
| $D$ | Diffusion rate of the signal | 874.5 | $\mu m^2/min$ | Estimated in this study |
| $x_0$ | Length of computational domain containing SNB10 in x-coordinate | 400 | µm | Estimated in this study |
| $y_0$ | Length of computational domain containing SNB10 in y-coordinate | 400 | µm | Estimated in this study |

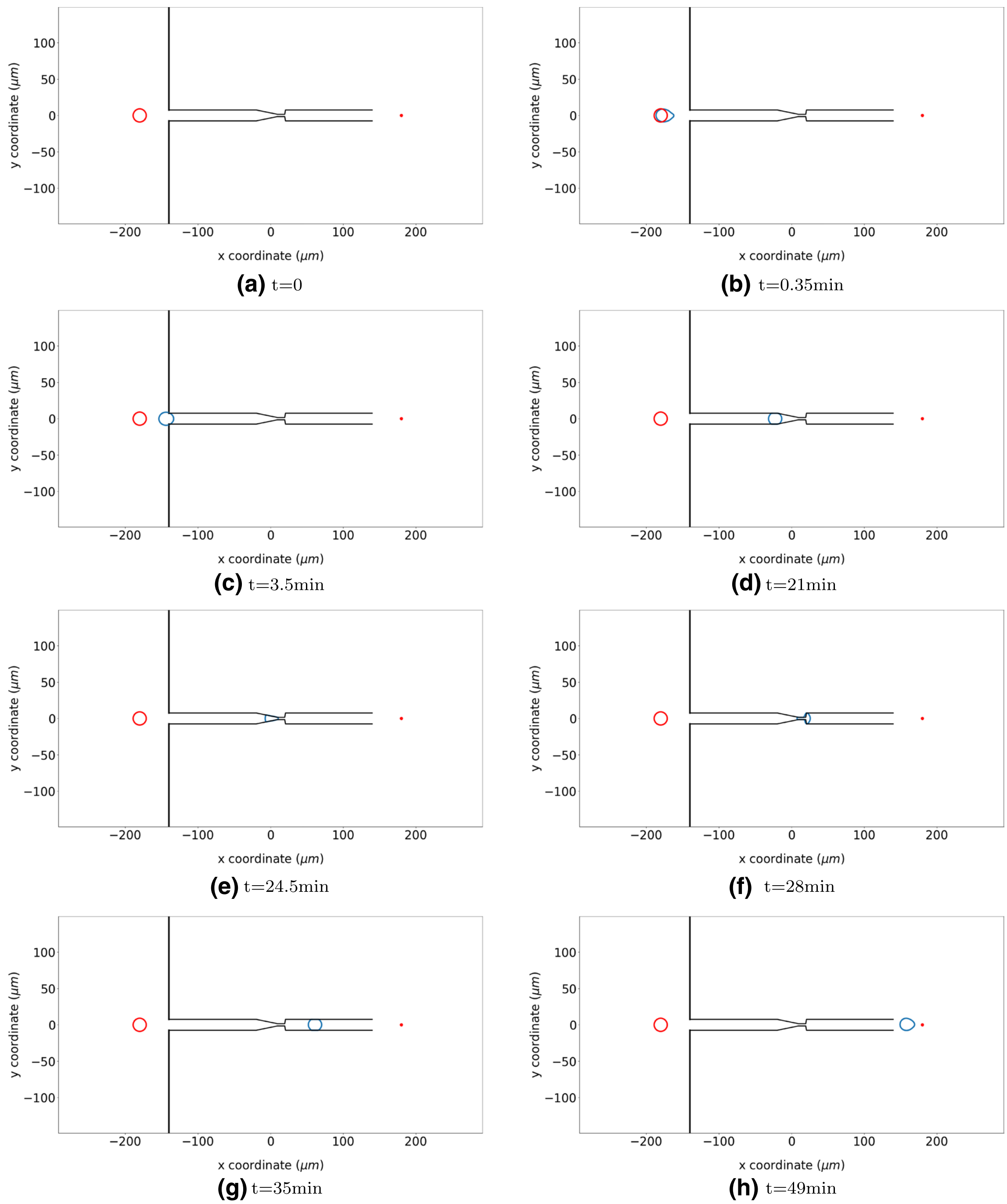

**Fig. 11** The figure shows the positions and shapes of cells at consecutive times by the blue contours when it travels through a microtube. The red circles show the original shape and position of cells, and the red dots in the end of the microtube is the point source of a signal

We run the simulations with four assumptions of the main mechanism provoking the active cell displacement: chemotaxis, fixed velocity with 10 μm/min, velocity generated from (6, 15) μm/min and (6, 20) μm/min in horizontal direction according to McCann et al. (2010). The number of samples and the Monte Carlo error of the occurrence of phase 3 collected from the Monte Carlo simulations of each aforementioned category are displayed in Table 6. Figure 13 illustrates the probability of the occurrence of phase 3, which is the stage when the cell stops monotonic forward migration. The mechanism that makes the cell move forward is not clear in Mak et al. (2013); hence, this could be reason for a mismatch between the experimental

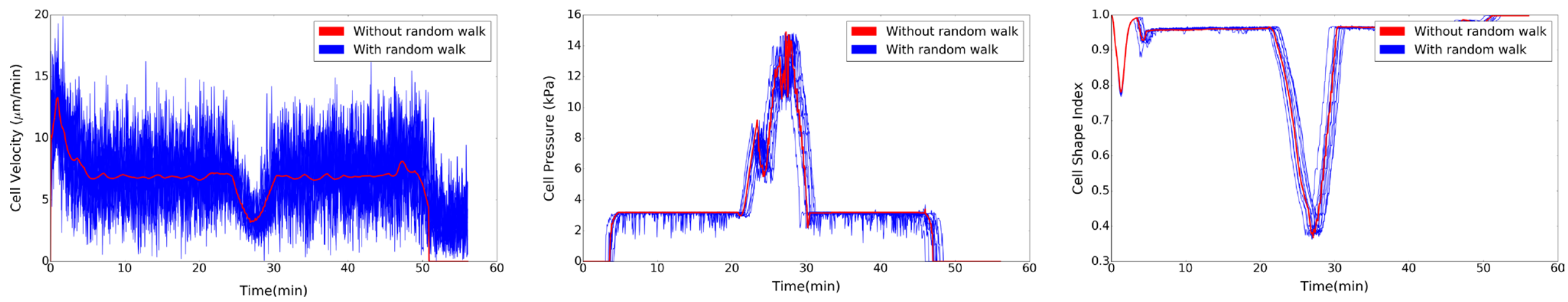

**(a)** Cell Velocity **(b)** Cell Pressure **(c)** Cell Shape Index

**Fig. 12** The cell velocity, pressure and shape index over time when the cell migrates through the microtube, where there is a 10 μm subnucleus barrier. The simulation mimics the experiment in Mak et al. (2013)

**Table 5** Parameter values used in the application of cell going through a microtube

| Parameter | Description | Distribution | Source |
|---|---|---|---|
| $\mu_f$ | Friction coefficient for the cell going through the microtube | $U(0.03, 0.06)$ | Angelini et al. (2012) |
| $\mu_m$ | The coefficient of cell mobility reduction | $U(0.6, 1)$ | Estimated in this study |

**Table 6** Monte Carlo simulations of various models, in which the main mechanisms of cell active displacement differ

| | Mechanism of cell active displacement | Number of samples from Monte Carlo simulations | Monte Carlo error of the occurrence of phase 3 |
|---|---|---|---|
| Simulation 1 | Chemotaxis | 1400 | $9.2171 \times 10^{-3}$ |
| Simulation 2 | Fixed velocity $v = 10$ | 1390 | $9.7022 \times 10^{-3}$ |
| Simulation 3 | Fixed velocity $v \in (6, 15)$ | 1360 | $1.2102 \times 10^{-2}$ |
| Simulation 4 | Fixed velocity $v \in (6, 20)$ | 1378 | $1.0621 \times 10^{-2}$ |

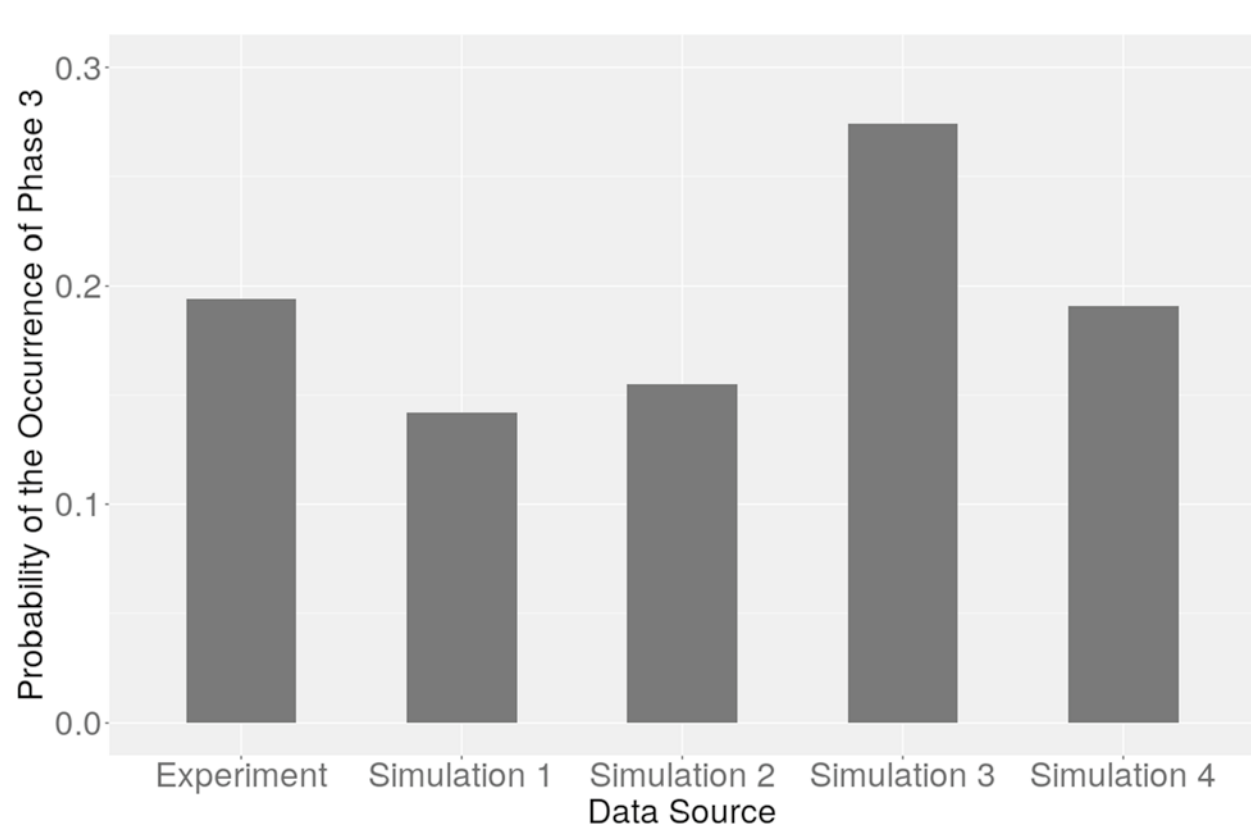

**Fig. 13** The probability of the occurrence phase 3 in Mak et al. (2013) and from Monte Carlo simulations by implementing different mechanisms of cell active displacement (see Table 6 for more information). The parameter values are taken from Table 5

and simulated results. However, the results are significantly close regarding the probability of the occurrence of phase 3. Coincidentally, according to our current simulations, with either chemotaxis or velocity 10 μm/min, the probability of the occurrence of phase 3 is the same in the first 3 digits.

Furthermore, the time cost of each phase is recorded and shown in Fig. 14. The results with different modelling settings or simulations do not show many differences, in particular between chemotaxis and fixed velocity $v = 10$. The reason is that the microtube restricts the displacement of the cell in the vertical direction; therefore, the cell mainly migrates in the horizontal direction. In general, the results between any simulation and the experiment differ more, compared with the results of the occurrence of phase 3, in particular, phase 1 and phase 3. Therefore, to investigate the possible reasons of mismatching results in phase 1 and phase 3 in phase time, we reran the simulation with the same settings as Simulation 4 in Table 6, except for the cell stiffness modified to $E_c = 1$. The results are shown in Fig. 15. With a softer cell, the simulation data in phase 3 match better with the experimental data. However, now a discrepancy between simulation and experiments results in phase 4 instead.

There are several possible reasons causing the discrepancy in the time interval of each phase. Firstly, for phase 3, we only obtain valid data when the cell moves non-monotonically, which results into a reduction of the sample size of the

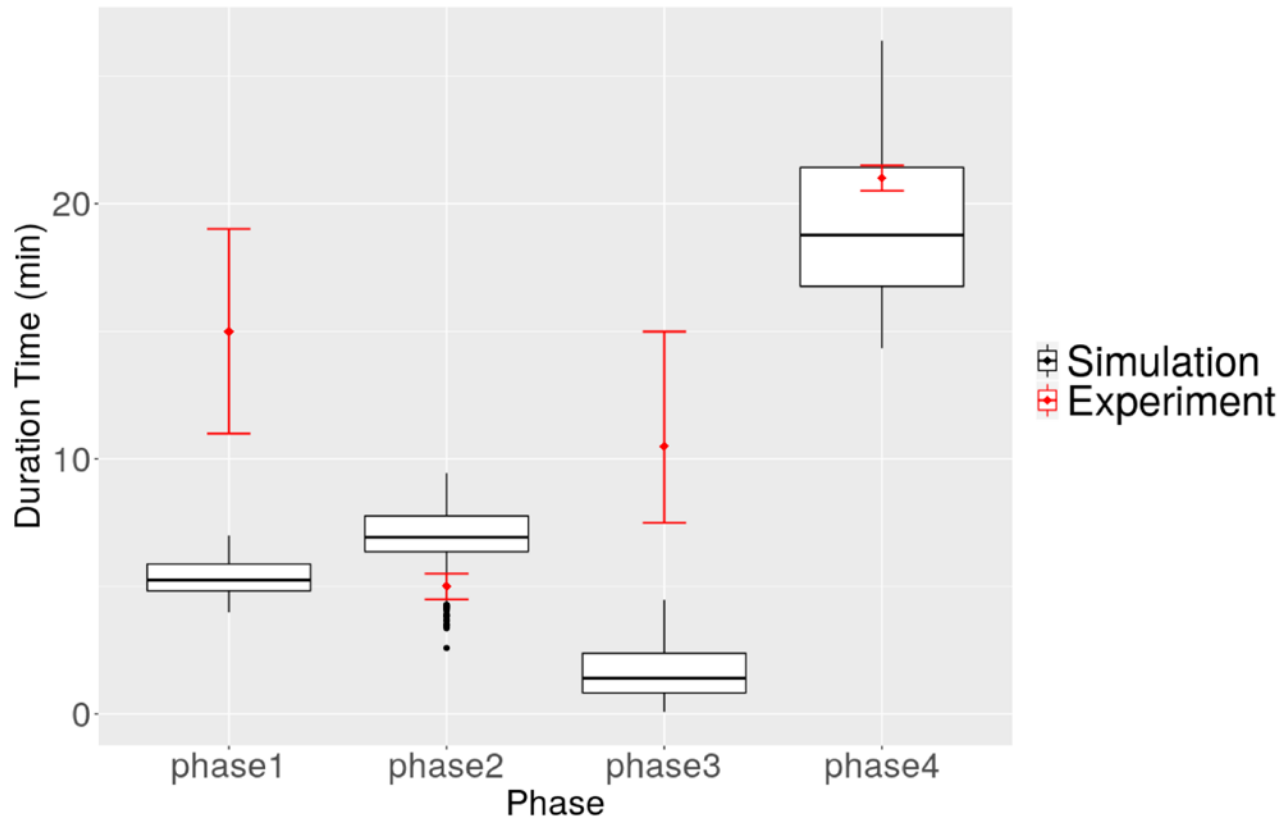

**(a)** Chemotaxis

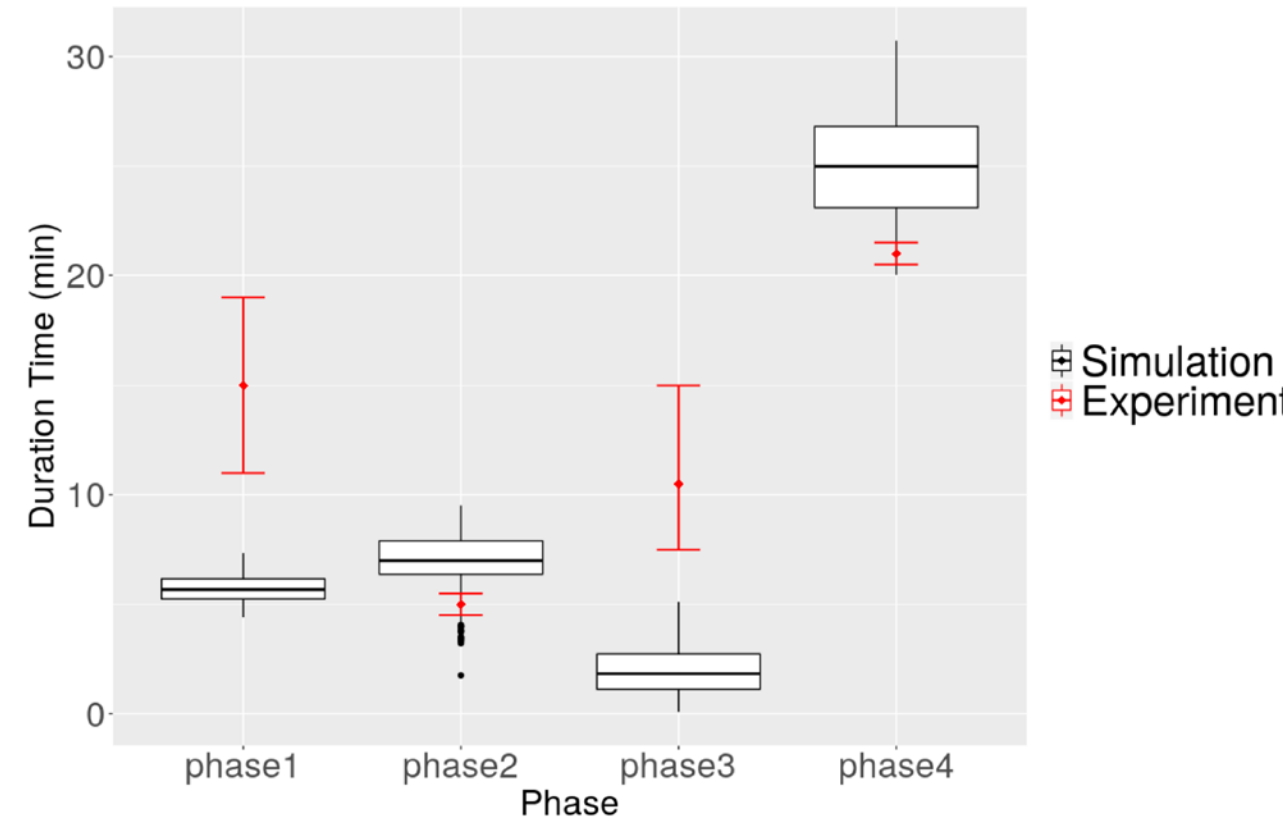

**(b)** Fixed velocity of $10\mu m/min$

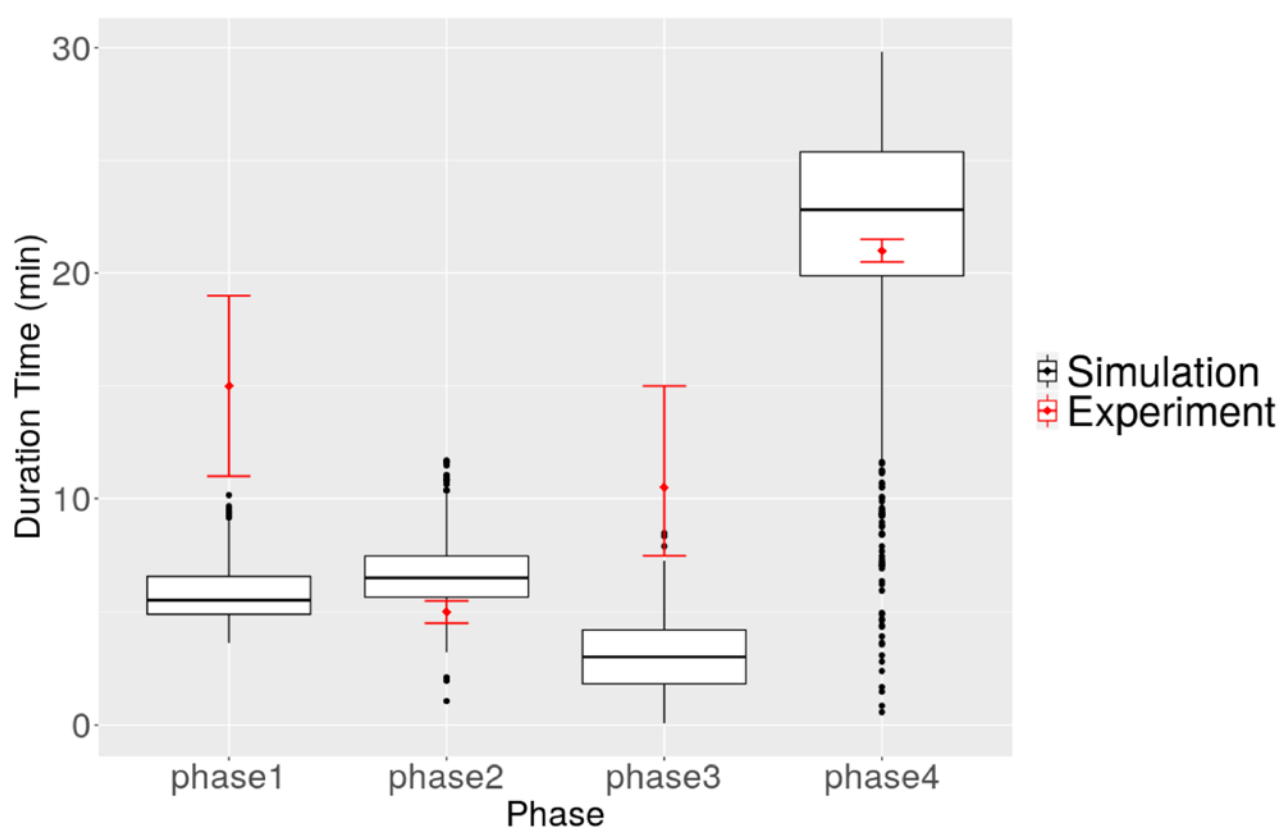

**(c)** Random velocity generated from $(6, 15)\mu m/min$

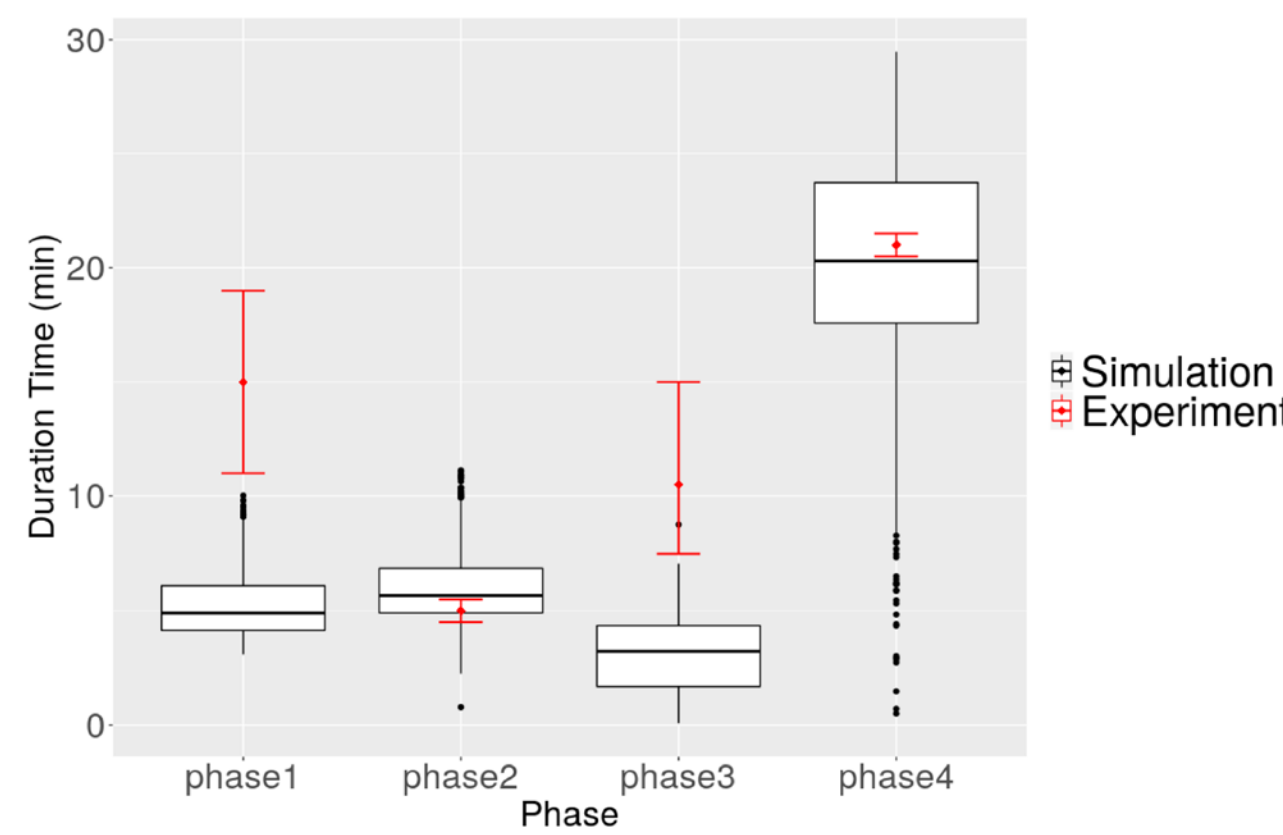

**(d)** Random velocity generated from $(6, 20)\mu m/min$

**Fig. 14** The figure shows the time cost of each phase in Mak et al. (2013) and from Monte Carlo simulations by using the model. Red dots with the error bar represent the experimental data from Mak et al. (2013) and the box plots are the data collected from the simulations

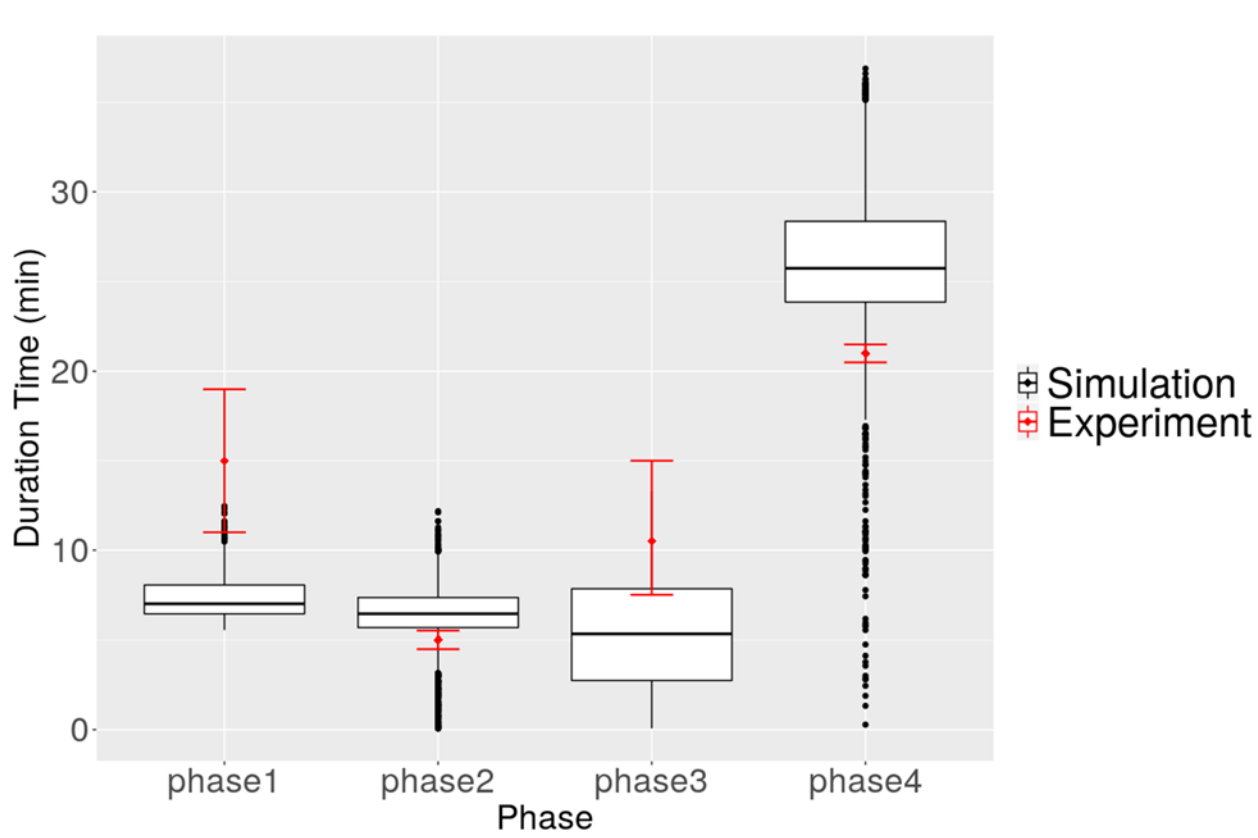

**Fig. 15** The time cost of each phase from the Monte Carlo simulation, in which the cell stiffness is $E_c = 1$ and the cell velocity is randomly generated from (6, 20). Red dots with error bar represent the experimental data from Mak et al. (2013) and the box plots are the data collected from the simulation

simulation data. Apparently the cell “dwells” and “doubts” whether it “wants” to keep on going if its pathway is (partially) obstructed. Secondly, the length of LC is not stated clearly in Mak et al. (2013); therefore, we could only estimate that from the scale in the figures. Thirdly, it is not clear if the velocity of active migration of the cell is constant, while in our simulation, the velocity can change over time, depending on the gradient of the chemotactic signal. Despite all these uncertainties, we still managed to reproduce the results which are close to the experimental results. Fourthly, the transaction of each phase from Mak et al. (2013) to our simulations may cause a mismatch of the duration time of each phase. Fifthly, it has been observed in Mak and Erickson (2013) that after the first time moving through the narrow channel, cells deform easier to move faster through the following narrow channels, which may indicate that the cell characteristics change regarding its geometry.

## 4 Conclusions

A phenomenological model for cell shape evolution and migration has been developed, with primary focus on the mechanics of the extracellular environment. Furthermore, the impact of passive convection, due to local displacements within the extracellular matrix, on the evolution of the cell shape has been taken into account. A morphoelastic model has been used in the current study to incorporate permanent deformations of the extracellular tissue. The model can be applied to mimic several microscopic biological observations such as cell deformation and migration during wound contraction and cancer metastasis. To validate the model, the experimental set-up in Mak et al. (2013) has been modelled. This experiment entailed cell migration through microtubes with different widths and with a varying width over the length. The model is able to reproduce the most important trends that were observed in the experimental data despite some experimental uncertainties such as the determination of which phase a cell is in during the transmigration process. Furthermore, the current model provides a basis that can be expanded to describe more experimentally observed phenomena in cell geometry.

**Acknowledgements** The authors appreciate China Scholarship Council (CSC) for the financial support for this project. The work was partially supported by the Israeli Ministry of Science and Technology (MOST) Medical Devices Program (Grant no. 3-17427 awarded to Prof. Daphne Weihs).

### Declarations

**Conflict of interest** The authors declare that they have no conflict of interest.